\documentclass[10pt, conference, letterpaper,anonymous]{IEEEtran}
\IEEEoverridecommandlockouts
\usepackage{cite}
\usepackage{amsmath,amssymb,amsfonts}
\usepackage{graphicx}
\usepackage{textcomp}
\usepackage{xcolor}
\usepackage{array}
\usepackage{booktabs}
\usepackage{setspace}

\usepackage{titlesec}
\usepackage{graphicx}
\usepackage{subcaption} 

\usepackage{algorithm}
\usepackage{algorithmicx}
\usepackage{algpseudocode}
\usepackage{amsmath}
\usepackage[top=2cm, bottom=2cm, left=2cm, right=2cm]{geometry}
\def\BibTeX{{\rm B\kern-.05em{\sc i\kern-.025em b}\kern-.08em
		T\kern-.1667em\lower.7ex\hbox{E}\kern-.125emX}}

\begin{document}
	\bibliographystyle{unsrt}
	\title{Energy Conserved Failure Detection for NS-IoT Systems\\
	
}

\author{
	\IEEEauthorblockN{Guojin~Liu\IEEEauthorrefmark{2},\,Jianhong~Zhou\IEEEauthorrefmark{1}\IEEEauthorrefmark{2}\IEEEauthorrefmark{3}, \emph{IEEE member},\,Hang~Su\IEEEauthorrefmark{2},\,Biaohong~Xiong\IEEEauthorrefmark{2},\,Xianhua~Niu\IEEEauthorrefmark{2}\IEEEauthorrefmark{3}}\\
	\IEEEauthorblockA{\IEEEauthorrefmark{2}School of Computer and Software Engineering, Xihua University\\
		\IEEEauthorrefmark{3}National Key Laboratory of Science and Technology on Communications,\\
		University of Electronic Science and Technology of China\\
		\IEEEauthorrefmark{1}Corresponding Author E-mail: zhoujh@uestc.edu.cn}\vspace {-4mm}\\
}

\maketitle

\begin{abstract}
	
Nowadays, network slicing (NS) technology has gained widespread adoption within Internet of Things (IoT) systems to meet diverse customized requirements. In the NS based IoT systems, the detection of equipment failures necessitates comprehensive equipment monitoring, which leads to significant resource utilization, particularly within large-scale IoT ecosystems. Thus, the imperative task of reducing failure rates while optimizing monitoring costs has emerged. In this paper, we propose a monitor application function (MAF) based dynamic dormancy monitoring mechanism for the novel NS-IoT system, which is based on a network data analysis function (NWDAF) framework defined in Rel-17. Within the NS-IoT system, all nodes are organized into groups, and multiple MAFs are deployed to monitor each group of nodes. We also propose a dormancy monitor mechanism to mitigate the monitoring energy consumption by placing the MAFs, which is monitoring non-failure devices, in a dormant state. We propose a reinforcement learning based PPO algorithm to guide the dynamic dormancy of MAFs. Simulation results demonstrate that our dynamic dormancy strategy maximizes energy conservation, while proposed algorithm outperforms alternatives in terms of efficiency and stability.
\end{abstract}


\section{Introduction}
The integration of  Network Slicing (NS) technology into Internet of Things (IoT) systems has garnered increasing attention. Through the incorporation of NS technology into IoT, operators can establish multiple dedicated, virtualized, and mutually isolated logical networks atop a shared physical infrastructure, which could effectively address various specialized demands\cite{alliance2016description}. In parallel with conventional IoT systems, the reliability of NS based IoT (NS-IoT) systems remains susceptible to device failures, which may potentially lead to adverse consequences such as network latency, bandwidth constraints, data loss, and security vulnerabilities\cite{9663233}. Thus, mitigating the failure rate of IoT devices within the NS-IoT systems has emerged as a pivotal concern. Conventionally, the prevalent approach involves deploying a monitoring mechanism to promptly detect failures by monitoring all devices, thereby reducing performance degradation attributed to such failures. However, handling post-failure exceptions may not always align with user Service Level Agreement (SLA) and could introduce security vulnerabilities. Preemptive exception prediction and avoidance are the preferred strategies.

Introducing Network Function Virtualization(NFV) technology and Network Data Analysis Function (NWDAF) into IoT systems ensures adherence to pre-defined system design and requirements, which can effectively mitigate the challenge of detecting device anomalies. The NWDAF is introduced by the 3rd Generation Partnership Project (3GPP) in Release 17 \cite{9832914}. It is mainly responsible for aggregating data from other Network Functions (NFs), training the data using embedded Machine Learning (ML) models, and subsequently transmitting the training outcomes to the respective NFs to discern abnormal conditions in IoT devices \cite{lei20215g}. NFV technology is leveraged to virtualize the anomaly detection functions into virtual functions, which are hosted through on server infrastructure. Subsequently, all data collected by these virtual functions can be transmitted to the NWDAF module residing on the server for rigorous analysis and processing. However, both monitoring and analysis processes consume critical network resources, including server computational capacity for data analysis, memory for storing data attributes, and network bandwidth for data transmission \cite{9537521}. When the data collection is extended to encompass a substantial volume of IoT devices, the consumption of these resources becomes increasingly critical. Consequently, the conventional approach of monitoring all IoT devices becomes unfeasible for forthcoming large scale IoT system.

In this paper, we presents an innovative dynamic dormancy monitoring mechanism in an NS-IoT system, which is grounded in the pre-existing NWDAF framework. It capitalizes on NFV technology to abstract monitoring functions for a substantial volume of devices into distinct Monitor Application Functions (MAFs). Each MAF is tasked with overseeing devices, which exhibit similar probabilities of encountering anomalies within a shared application context. A dormant MAF signifies that the devices under its supervision are not required to transmit data for analysis, thereby enabling resource conservation. The main contributions of this study can be succinctly summarized as follows.

1) We integrate the NWDAF framework into existing IoT systems to deploy a novel NS-IoT system, and leverage NFV technology to logically abstract the devices monitoring functions into Monitoring Application Functions (MAFs).

2) Within the aforementioned NS-IoT novel systems, we introduce a dynamic dormancy monitoring mechanism for MAFs. This mechanism dynamically adjusts the sleep duration of various MAFs to regulate when devices suspend data transmission for analysis during designated sleep intervals.

3) We develop a Reinforcement Learning (RL)-based approach to obtain an optimized strategy for learning device anomaly patterns, thus diminishing monitoring energy consumption while upholding the requisite detection accuracy. 

The rest of the paper is structured as follows: In Section II, the system model of the NS-IoT system and the dynamic dormancy monitoring mechanism are introduced. Section III formulates the problem of energy conservation as an optimization problem. In Section IV, the problem is summarized as an MDP process and the algorithm flow is described in detail. Section V simulates the proposed monitoring sleep strategy and discusses the experimental results. Section VI provides the conclusion of this paper.

\section{NS-IOT System Model}
The MAF-NWDAF based NS-IoT system model is illustrated in Fig. 1. First, IoT devices are categorized into device clusters within each NS, based on their data transmission characteristics. This entails grouping devices of the same type under each NS, all of which are overseen by a corresponding MAF. Devices within each NS are required to transmit data to the associated MAF via the upload link for subsequent analysis. Notably, an MAF cannot directly establish a subscription with NWDAF from the server;  instead, it relies on the Network Exposure Function (NEF) as an intermediary interface The NEF is tasked with facilitating NWDAF service provision to the MAF and forwarding analysis outcomes back to the MAF. Functioning as a vital data analysis module within the network,  NWDAF shoulders the primary responsibilities of data collection, processing, analysis, and the generation of network data analysis reports. It serves as a key enabler of data analysis services and extends support to the MAFs.

\begin{figure}[!t]
	\centering
	\captionsetup{font=footnotesize, labelsep=period}
		\includegraphics[width=3.3in]{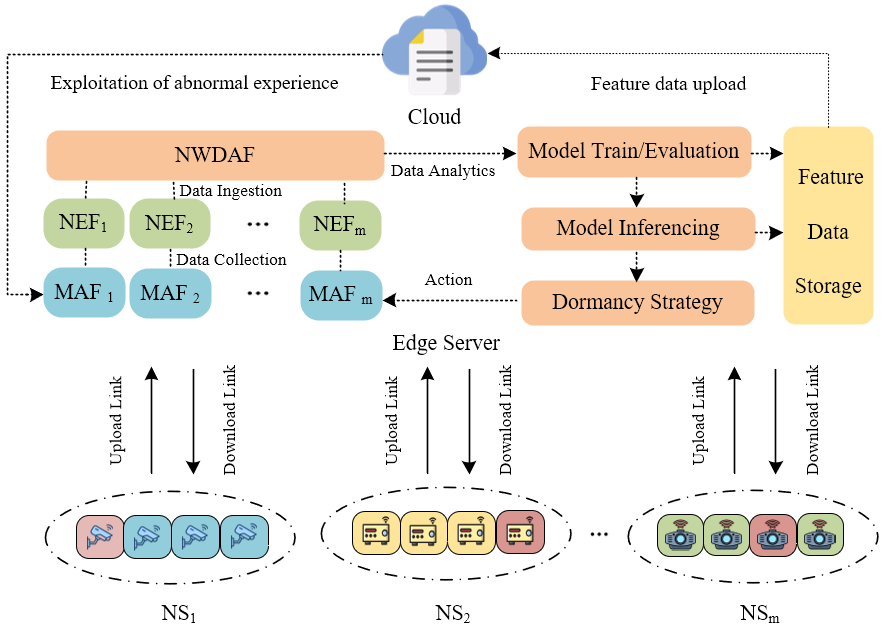}
	\caption{MAF-NWDAF based NS-IoT system model.}
	\label{fig1}
\end{figure}

The machine learning functionality deployed within NWDAF can be delineated into two distinct logical functions: train and analysis. The training function assumes responsibility for model training, encompassing both initial training and periodic retraining using new training data to maintain model currency. The analysis function employs the trained model \cite{9786853} to generate results for both the MAF and the server. For the server, subsequent to model training by  NWDAF,  data is analyzed to extract essential feature information, which is archived as historical experience for future reference. This repository proves invaluable in expeditiously addressing analogous issues should they arise with other IoT devices. Concurrently,  for the MAF, leveraging the trained model, NWDAF conducts analyses to identify devices exhibiting potential faults that warrant monitoring.  Subsequently, NWDA determines the monitoring duration and sleeping intervals for each MAF, disseminating this information to the respective MAF through NEF. During MAF monitoring periods,  IoT devices (IoTDs) within the corresponding cluster are tasked with transmitting data to the server. In contrast, during MAF dormancy phases, a signal relayed to the IoTDs in the corresponding cluster to halt data transmission until the MAF resumes monitoring. Obviously, substantial energy conservation can be achieved by suspending data transmission and halting NWDAF computations during dormancy. Our goal is to curtail energy consumption through the employment of the MAF dormancy mechanism. Next, we will model the energy consumption for encompassing four distinct process: device-server transmission, server processing, server-cloud transmission, and abnormal behavior.
\subsection{Device-Server Transmission Energy Consumption}
Assume that the system includes $ M $ network slices, and each network slice includes $ N $ IoTDs.  The NWDAF feeds back results to all MAFs simultaneously. The time interval between any two feedback results is named as decision-making round and the duration is $ T $, which is composed of monitoring duration $ t_1 $ and dormancy duration $ t_2 $. Please note that $ T $ is the same for all MAFs and does not change with time. We assume that at the beginning of the $ k_{th} $ decision-making round, the NWDAF generate and feedback the monitoring duration  $ t_{1,m}^k $ and the dormancy duration $ t_{2,m}^k $ to the $ MAF_m $. The transmission energy consumption of the $ m_{th} $ NS for the $ k_{th} $ decision-making round can be denoted as

\begin{equation}
	E_1^{tran}\left ( k \right ) =\sum_{m=1}^{M} \sum_{n=1}^{N}p_{m,n}t_{1,m}^k,
\end{equation}
where $ p_{m,n} $ is the transmission power of the $IoTD_{m,n}$.

\subsection{Server Processing Energy Consumption}
We assume that the power consumption per unit time required by the server for data processing is $e^{deal}$, and the given computing resources of the server is $ P $. Since the calculation frequency $ c $ of the server (the calculation frequency of the CPU per second) is constant, the calculation capability $ F $ of the server can be expressed as $F^{deal}=cP$. Meanwhile, due to the different types and priorities of the NSs, the transmission rates between the devices and the corresponding MAF in different NSs are also different. The higher the NS priority, the faster the transmission rate and the greater the network bandwidth required. Thus, the transmission rate $ v_{m,n} $ for the $ IoTD_{m,n} $ can be expressed as:

\begin{equation}
	v_{m,n}=B_mlog_2\left ( 1+\frac{p_{m,n}}{N_m}  \right ) ,
\end{equation}
where $ N_m $ is the modeled noise between the $ IoTD_{m,n} $ and  the $ MAF_m $. Thus, during the $ k_{th} $ decision-making round, the total data transmitted to the server $ D_{m,n}^k =v_{m,n} t_{1,m} ^k $. The server processing energy consumption $E_1^{deal}\left ( k \right ) $ can be expressed as:

\begin{equation}
	\begin{aligned}
		E_1^{deal}\left ( k \right )  &= e^{deal} \frac{\sum_{m=1}^{M}\sum_{n=1}^{N}D_{m,n}^k}{cP} \\
		&= e^{deal} \frac{\sum_{m=1}^{M}\sum_{n=1}^{N}v_{m,n}t_{1,m}^k}{cP}.
	\end{aligned}
\end{equation}

\subsection{Server-Cloud Upload Energy Consumption}
Under the dormancy mechanism, the server will retain the data containing exceptions and use them as experience when devices encounter similar anomalies. Specifically, after NWDAF sends the determined monitoring duration and dormancy duration to the $MAF_m$, the server will store all data received during the monitoring period, which could help generate an exception report and help quickly determine the type of exception. Meanwhile, it could facilitate quick processing in case of similar anomalies in the future, thereby saving the time and costs for exceptions recovery. However, the cumulative storage of the data over time may lead to insufficient memory space in the server. In such cases, the server needs to upload data to the cloud, which results in server-cloud upload energy consumption $E_1^{up}$. Assume that the power consumption required for the server to upload a unit memory block of monitoring data to the cloud is $e^{up}$. During the $ k_{th} $ decision-making round, the energy consumption $E_1^{up}\left ( k \right ) $ to upload data from server to cloud can be represented as:

\begin{equation}
	E_1^{up}\left ( k \right ) =e^{up}\frac{\beta S^{size}}{S^{unit}} f,  
\end{equation}
where $S^{size}$ indicates the size of the server's memory. $ \beta $ is the storage space occupancy threshold. It means that when ratio of actual data storage space to the total storage space exceeds this threshold, the data should be uploaded. $ f $ indicates the upload frequency, and $S^{unit}$ indicates the size of unit memory block. Since $e^{up}$, $S^{size}$, and $S^{unit}$ are constants, the upload energy consumption is only related to the upload frequency $ f $, which depends on the total amount of data transmitted to the server from the $IoTD_{m,n}$. Thus, the upload frequency $ f\left ( k \right )  $ of the $ k_{th} $ decision-making round can be expressed as:
\begin{equation}
	f\left ( k \right ) =\frac{\sum_{m=1}^{M}\sum_{n=1}^{N}D_{m,n}^k}{\beta S^{size}} .
\end{equation}

According to (4) and (5), the relationship between upload energy consumption and data volume can be expressed as:

\begin{equation}
	\begin{aligned}
		E_1^{up}\left ( k \right )  &= e^{up}\frac{\sum_{m=1}^{M}\sum_{n=1}^{N}D_{m,n}^k}{S^{unit}}\\
		&= e^{up}\frac{\sum_{m=1}^{M}\sum_{n=1}^{N}v_{m,n}t_{1,m}^k}{S^{unit}} .
	\end{aligned}
\end{equation}
\subsection{Abnormal Energy Consumption of Equipment}
If a device abnormality occurs during the monitoring duration, it can be detected and recovered immediately. Thus, no abnormal operation involved and there is no abnormal energy consumption. However, if the device abnormality occurs during the dormancy duration, it can only be detected and recovered when the next monitoring duration starts. It means that abnormal operation of the device will continue for a while, which results in abnormal energy consumption. The reason is that the abnormal devices may continuously record abnormal states, collect data more frequently, and eventually generate error reports for servers. Assume that all devices under the same slice have the same probability of abnormality, and the abnormality occurrence  of one device is uniformly distributed with probability $ \frac{1}{T^{\prime} }   $ . Thus, the time interval of the abnormality occurrence for any device is $ T^{\prime} $. $ k $ represents the number of the decision-making rounds for NWDAF. Thus, for the network slice $ m $, the exception duration of one device $ t_{3,m}^k  $ during the MAF sleeping period in the $ k_{th} $ decision-making round can be expressed as:

\begin{equation}
	t_{3,m}^k=kT-\left ( H_k+1 \right ) T^{\prime}  ,   
\end{equation}
where $ H_k $ represents the actual cumulative number of exceptions before the end of the $ k_th $ monitoring duration. For analysis convenience, we assume that the abnormal power consumption of $ IoTD_{m,n} $ during the dormancy period is  $ e_{m,n}^l $ and the number of abnormal devices in the kth decision-making round is $ I \left ( I\le N \right )  $. Then the abnormal energy consumption $ E_2 $ of the equipment can be expressed as:
\begin{equation}
	\small
	E_2^k=\left\{\begin{array}{lc}
		0 \quad {\text{if}  \left ( k+\alpha  \right )T< \left ( H_k+1 \right )T^{\prime }<  \left ( k+\alpha  \right )T+t_{1,m}^{k+\alpha +1}  }\\
		\sum_{m=1}^{M}\sum_{i=1}^{I}e_{m,n}^lt_{3,m}^ki   \quad \text{else}.
	\end{array} \right.
\end{equation}

Among them, $ \alpha =0,1,2, \cdots  $, represents the number of rounds after the $ k_{th} $ decision-making round. The condition $ \left ( k+\alpha  \right )T< \left ( H_k+1 \right )T^{\prime }<  \left ( k+\alpha  \right )T+t_{1,m}^{k+\alpha +1} $ means that when the  abnormality of $ IOTD_{m,n} $ happens during the monitoring period of the $ MAF_m $, the abnormality could be detected and recovered in time and the abnormal energy consumption $ E_2^k $ is 0.
\section{Problem Description}
Based on the principle of the above mentioned dormancy mechanism, the settings of $ t_{1,m}^k $ and $ t_{2,m}^k $ may seriously affect the system energy consumption. As different types of equipment have different probabilities of abnormality. If no abnormality occurs during continuous monitoring duration, the monitoring process is ineffective and may result in energy consumption $ E_1^k $, which includes Device-Server transmission energy consumption $E_1^{tran}\left ( k \right ) $, server processing energy consumption $ E_1^{deal}\left ( k \right ) $ , and Server-Cloud upload energy consumption $ E_1^{up}\left ( k \right ) $ . On the other hand, if abnormality occurs during the dormancy duration, the abnormal operation will continue until next monitoring duration begins. Thus, the resulting abnormal operation energy consumption $ E_2^k $ is ineffective for the system operation and should be reduced. The longer the duration of abnormal device states, and the more anomalies generated within the NS, the greater the energy consumption required. Meanwhile, compared with other energy consumption, the energy consumption of MAF from sleep to restart is very small and can be ignored.  Thus, the total invalid energy consumption of the $ k_{th} $ decision-making round $ E_{total}^k $ can be represented as:
\begin{equation}
	E_{total}=E_1^{tran}\left ( k \right ) +E_1^{deal}\left ( k \right )+E_1^{up}\left ( k \right )+E_2^k. 
\end{equation}

Our goal is to obtain an optimal dormancy strategy to guide the sleeping and monitoring process for $MAF_m$, so as to save total invalid energy consumption for long-term, which includes ineffective monitoring and abnormal device operation energy consumption. The problem can be formulated as follows:

\begin{equation}
	\begin{aligned}
		P_1.\quad\quad&\underset{t_{1,m}^k}{min} \sum_{k=1}^{K}E_{total}^k \\
		s.t.\quad\quad&C1:0\le i\le N,\forall i\\
		&C2:1\le t_{1,m}^k\le T,\forall t_{1,m}^k\\
		&C3:\tau \le \frac{\sum_{k=1}^{K}e_{mo}^k}{\sum_{k=1}^{K}e^k }\le 1.
	\end{aligned}
\end{equation}

Where $ K $ represents the total number of decision-making durations received by MAF from NWDAF. $ e_{mo}^k $  and $ e^k $ are at the $ k_{th} $ decision-making round, the cumulative number of exceptions occuring during the monitoring duration and the cumulative number of exceptions occurring during the whole round, respectively.

C1 means that the number of abnormal devices cannot exceed the number of the devices monitored by the $MAF_m$. C2 means that the monitoring duration determined by NWDAF cannot exceed the decision-making duration. C3 is the constraint of monitoring accuracy, which should be greater than the required monitoring accuracy $\tau $. 

\section{Algorithm Design to Obtain the MAF Dormancy Strategy}

To solve the problem $ P1 $, NWDAF should  accurately predict the future abnormal patterns of devices by analyzing historical data, reduce the occurrence of invalid monitoring and detect the equipment abnormalities in time through dynamic adjustment of monitoring and sleep duration. However, the problem $ P1 $ is a typical NP-hard problem,  which cannot be solved in polynomial time.

Deep Reinforcement Learning (DRL) algorithms have a significant advantage in solving such problems. Thus next, we map the dynamic selection process of MAF monitoring and sleep durations, into a Markov Decision Process (MDP), and adopt the Proximal Policy Optimization (PPO) algorithm, which is a kind of DRL algorithm to solve the problem $ P1 $.
\subsection{MDP}\label{AA}

MDP is usually described as a five-tuple: $M=[S,A,P,R,\gamma ]$, which includes state space State, action space Action, and state transition probability $ P $ , reward $ R $ and decay factor $\gamma$. The dynamic dormancy process of MAF can be modeled as an MDP as follows.

Agent: It's the NWDAF, which receives data from all MAFs, analyzes to generate the dormancy strategy for all MAFs. It dynamically coordinates the monitoring and sleep durations for each MAF to minimize the energy required for monitoring while ensuring monitoring accuracy.

State: For NWDAF, all devices under monitored by different MAFs are independent, which means that we only need to ensure that the total ineffective energy consumption of devices under each MAF is minimal. Thus, we only need to analyze the energy-saving strategy for one MAF and then generalize the strategy to all MAFs. Thus, we define the duration of abnormal behavior of all devices during the $ (k-1)_{th} $ decision-making round as the state $\mathbf{s_k}$ for the $ k_{th} $ decision-making round, which can be expressed as :
\begin{equation}
	\mathbf{s_k} =\left \{ t_3^{k,1},t_3^{k,2},\dots ,t_3^{k,n} \right \}, 
\end{equation}
where $ \mathbf{s_k} $ is an one-dimensional vector. It is composed of $ n $ elements,  which represents the abnormal status of all IoTDs under the network slice $ M $. If no abnormality occurs to the $ IoTD_n $ during the dormancy duration of the $ (k-1)_{th} $ decision-making round, the status value $ t_3^{k,n}=0 $. 

Action: Since the length of each decision is constant, the dormancy duration will be known after the monitoring duration is obtained, the action $\mathbf{a_k}$ can be expressed as the monitoring duration sent by NWDAF to different MAFs, that is $ \mathbf{a_k} =\left \{ t_{1,1}^k,t_{1,2}^k,\dots ,t_{1,m}^k \right \}  $.

Reward: Our goal is to minimize additional energy consumption and need to maximize rewards. Therefore, the Reward within each decision time can be expressed as:

\begin{equation}
	R_k=\frac{1}{E_1^k+E_2^k} .
\end{equation} 

If $ MAF_m $ detects an abnormality during the monitoring period in the $ k_{th} $ decision-making round, the abnormal energy consumption  $ E_2^k $ is 0, then the system energy consumption in this round is only $ E_1^k $. Otherwise, the energy consumption includes $ E_1^k $ and $ E_2^k $.

\subsection{Design of PPO Algorithm under Dormancy Strategy}

It can be seen from the above MDP that the action of each $MAF_m$ is continuous, and there are currently many DRL algorithms for solving continuous action problems, literature \cite{schulman2017proximal} has proposed that PPO is a good algorithm way to solve the above types of problems. Therefore, we have designed an algorithm based on PPO-Clip for MAF-NWDAF sleep control, and its algorithm model is shown in Fig. 2. The main interaction process includes the environment, storage buffer, and network updates. 

\begin{figure}[!t]
	\centering
	\captionsetup{font=footnotesize, labelsep=period}
	\includegraphics[width=3.4in]{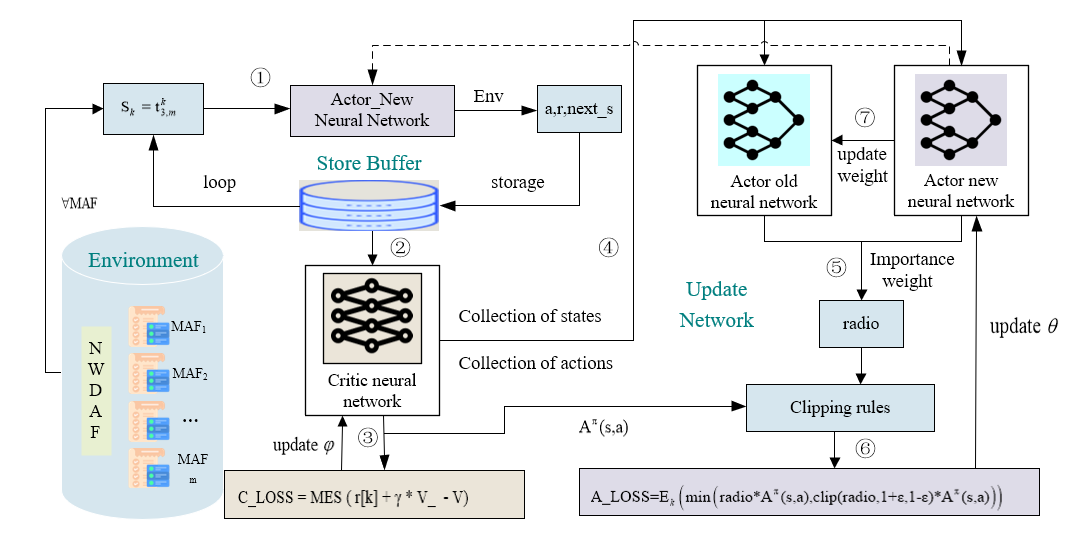}
	\caption{MAF sleep algorithm model diagram based on PPO-Clip.}
	\label{fig2}
\end{figure}

We update the parameters $\varphi $ of the value network by minimizing the mean squared error regression to fit the value function. Then, we compute the advantage function and use a clip-based method to update the policy network parameters $\theta $. The minimization optimization objective function is as follows:

\begin{equation}
	\begin{split}
		&J_{\theta_r}^{clip}(\theta) = E_{\pi_{\theta}}\left[ \min\left[ \frac{\pi_\theta(a_k|s_k)}{\pi_{\theta_{r}}(a_k|s_k)} A^{\pi_{\theta_r}}(s_k,a_k), \right. \right. \\
		&\left. \left. \text{clip}\left[ \frac{\pi_\theta(a_k|s_k)}{\pi_{\theta_r}(a_k|s_k)}, 1-\varepsilon , 1+\varepsilon  \right] A^{\pi_{\theta_r}}(s_k,a_k) \right] \right].
	\end{split}
\end{equation}

The usual update method is Adam stochastic gradient ascent \cite{sadiki2023deep}. The relevant algorithm is described in Algorithm 1.
\renewcommand{\algorithmicrequire}{\textbf{Input:}}  
\renewcommand{\algorithmicensure}{\textbf{Output:}} 

\begin{algorithm}[h]
	\caption{PPO-clip Algorithm Based on MAF Dormancy Strategy} 
	\begin{algorithmic}[1]
		\Require
		Initialize $lr$:  learning rate;
		$\gamma$: discount factor;
		$\theta$: parameters of the policy network;
		$\varphi$: parameters of the value network
		\Ensure
		MAF optimal dormancy strategy model 
		\For{$r = 0,1,2,\cdots$}
		\State According to $\pi \left ( \theta _r \right ) $ interacting with $MAF_m$, collect the trajectory data set $D_r =\left \{ {\tau _r} \right \} $.
		\State Calculate $V_{\varphi_r}\left ( s_k \right )$ of the current trajectory through $\varphi$, and then calculate the advantage estimate of the trajectory $A^{\pi _\theta {_r}} (s_{k} ,a_{k} ) $ with $ R_k $. Then update $\varphi$:
		\begin{equation*}
			\label{eq1}
			\varphi_{r+1} = \underset{\varphi}{\text{argmin}} \frac{1}{|D_r|K} \sum_{\tau \in D_r} \sum_{k=1}^{K} (V_{\varphi_r}(s_k) - \hat{R_k})^2			
		\end{equation*}
		\State For the objective function $J_{\theta_r }^{clip} (\theta )$, and use the PPO-clip optimization algorithm to update $\theta$:
		\begin{equation*}
			\label{eq1}
			\theta _{r+1}=\underset{\theta }{argmax} J_{\theta_r }^{clip} (\theta )
		\end{equation*}
	
		\EndFor
	\end{algorithmic}
\end{algorithm}

\section{Numerical Simulation Analysis}
In this section, we use the Python-based Pytorch framework to simulate the dormancy control of MAF by NWDAF. For NWDAF, different MAFs correspond to different monitoring device clusters, which is mainly reflected in the different device types, data types, and data transmission rates. We only need to simulate the process of one device cluster monitored by an MAF, which could be generalized to the other MAFs.

In the monitoring system, the action space includes the length of sleep and monitoring time, which is a continuous value, so we use the DDPG algorithm that also handles the continuous action space. For the DQN algorithm, although it is an algorithm that deals with discrete state spaces, if we fix the time unit length to 1 minute, and then set the sleep and monitoring time lengths to multiples of the unit length, convert it into multiple discrete actions, Thus next, we first compare the performance of the PPO, DDPG, and DQN algorithm, and use them to obtain the MAF dormancy strategy, which are named as PPO based strategy, DDPG based strategy, and DQN based strategy, respectively. We simulate the performance of the PPO based dormancy strategy and compare it with the DDPG based and DQN based strategies from three aspects such as system resource overhead, sleep duration, and monitoring accuracy. 

\subsection{Parameter Setting}
Monitoring system environment settings: Assume that IoTD faults are mapped to abnormal data, that is to say, some abnormal data are included while generating normal data, and the arrival of these abnormal data is subject to generation every 20-time units (the time unit is 1min). We simulate the scene when $AMF_m=1$. All parameter settings can be referred in table I.

\begin{table}[htbp]
	\centering
	\caption{EXPERIMENTAL PARAMETER SETTINGS}
	\small 
	\begin{tabular}{|c|c|c|c|}
		\hline
		\textbf{Parameter} & \textbf{Value} & \textbf{Parameter} & \textbf{Value} \\
		\hline
		$t$ & 20 min & $p_1$ & 0.5 kw/min \\
		\hline
		$MAF_m$ & 1 batch & $p_2$ & 1 kw/min \\
		\hline
		$n$ & 100 unit & $v$ & 10 MBps \\
		\hline
		$i$ & 10 unit & $c$ & 1 Hz \\
		\hline
		$P$ & 10 G/round & $e_{up}$ & 1 kw \\
		\hline
		$S^{unit}$ & 100 block & $\tau $ & 93\% \\
		\hline
	\end{tabular}
	\label{tab1}
\end{table}

\subsection{Simulation Results}
We set the number of decision-making rounds $ k $ to 15,000 rounds, and define 10 rounds of decision-making as one episode. We choose two typical RL algorithms, namely DDPG and DQN, as the comparison algorithms to compare the convergence speed and stability. The comparison result is shown as in Fig. 3(a). In terms of convergence speed, PPO reaches the maximum reward at around episode 3,00, while the other two algorithms converge significantly slower than PPO. The reason is that the PPO algorithm uses a clip optimization strategy, meanwhile, PPO employs important sampling techniques when updating policies. The stability of the algorithm is mainly reflected in the fluctuation range of Reward after convergence. The smaller the fluctuation, the more stable the algorithm is. It can be seen that the PPO algorithm is significantly better than the DDPG and DQN algorithms. Fig. 3(b) compares the system energy consumption of the four strategies. Average energy consumption per minute was calculated over the duration of 1,500 episodes. The full monitoring strategy always has the most ineffective energy consumption, with the average energy consumption being about 230kw. The DQN based strategy decays slowly and the average energy consumption is around 75kw, although DDPG is more efficient than DQN. However, due to its instability, it consumes too much energy. The average energy consumption of all 1500 episodes is about 49kw. The PPO based strategy produces the least system energy consumption of about 35kw, thanks to its excellent convergence performance.

\begin{figure}[!t]
	\centering
	\captionsetup{font=footnotesize, labelsep=period}
	\begin{subfigure}{0.48\linewidth} 
		\includegraphics[width=\linewidth]{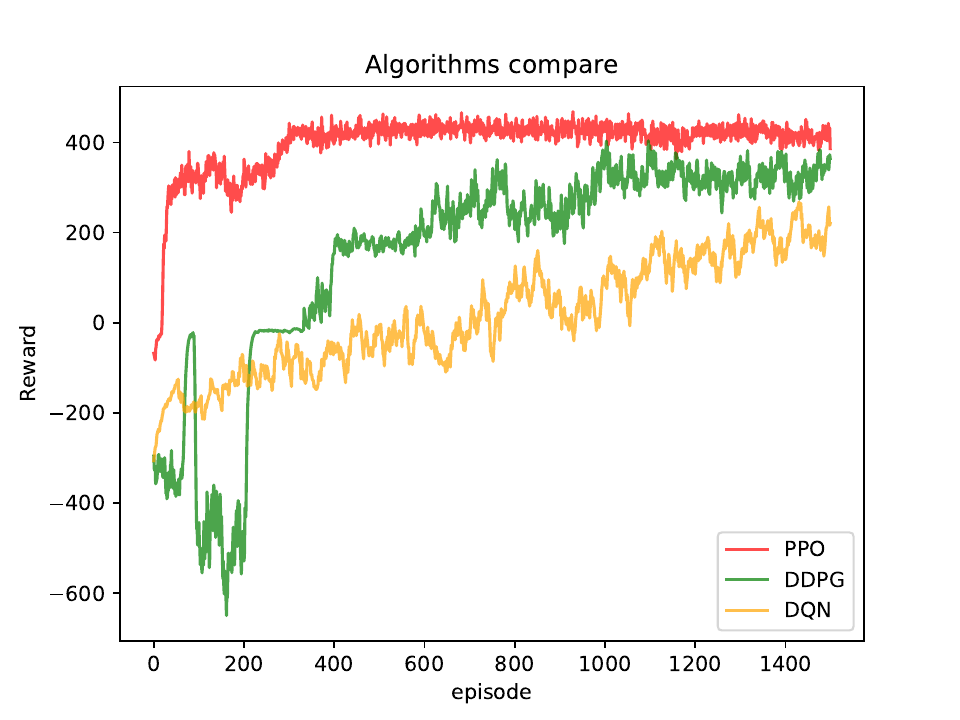}
		\caption{Comparison of convergence speed and stability among three algorithms.}
	\end{subfigure}
	\begin{subfigure}{0.46\linewidth} 
		\includegraphics[width=\linewidth]{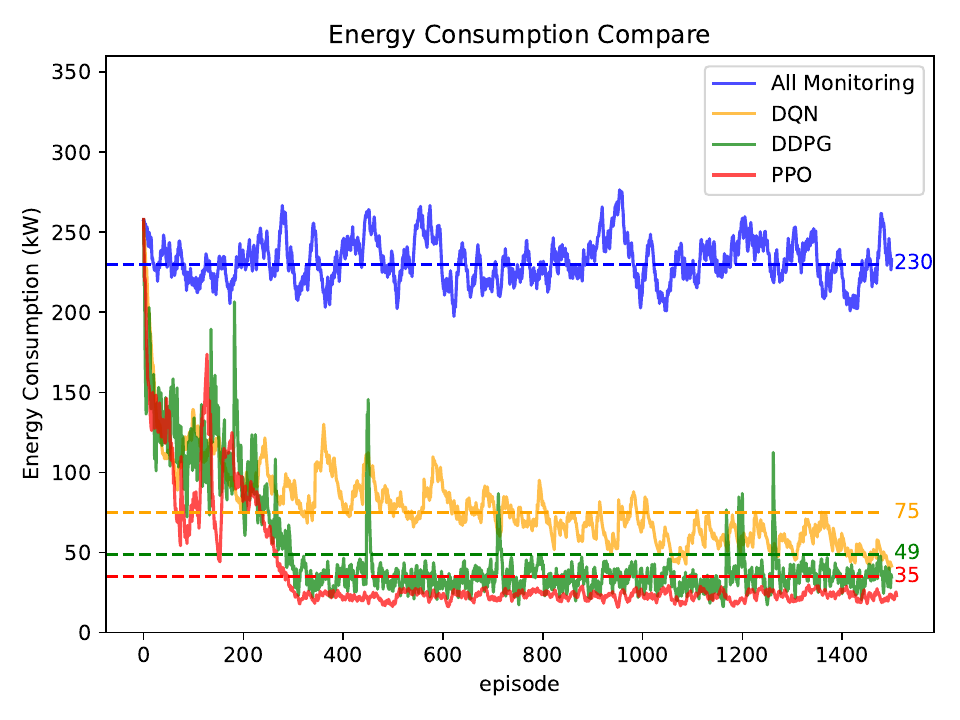}
		\caption{The energy consumption overhead of the system using different methods}
	\end{subfigure}
	\begin{subfigure}{0.48\linewidth} 
		\includegraphics[width=\linewidth]{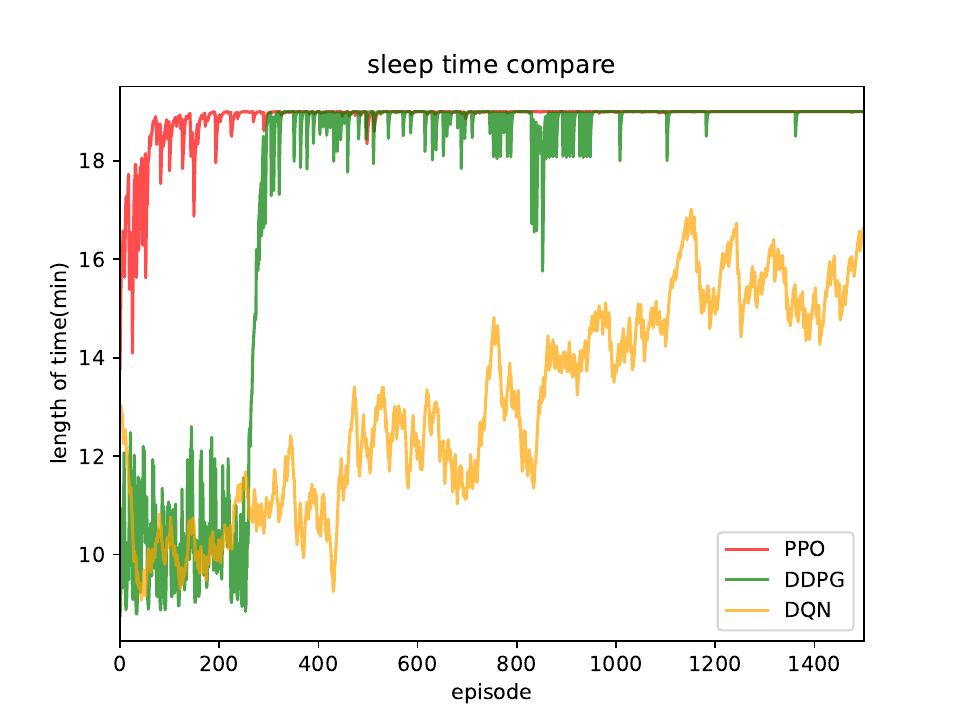}
		\caption{Comparison of sleep time length among three algorithms}
	\end{subfigure}
	\begin{subfigure}{0.48\linewidth} 
		\includegraphics[width=\linewidth]{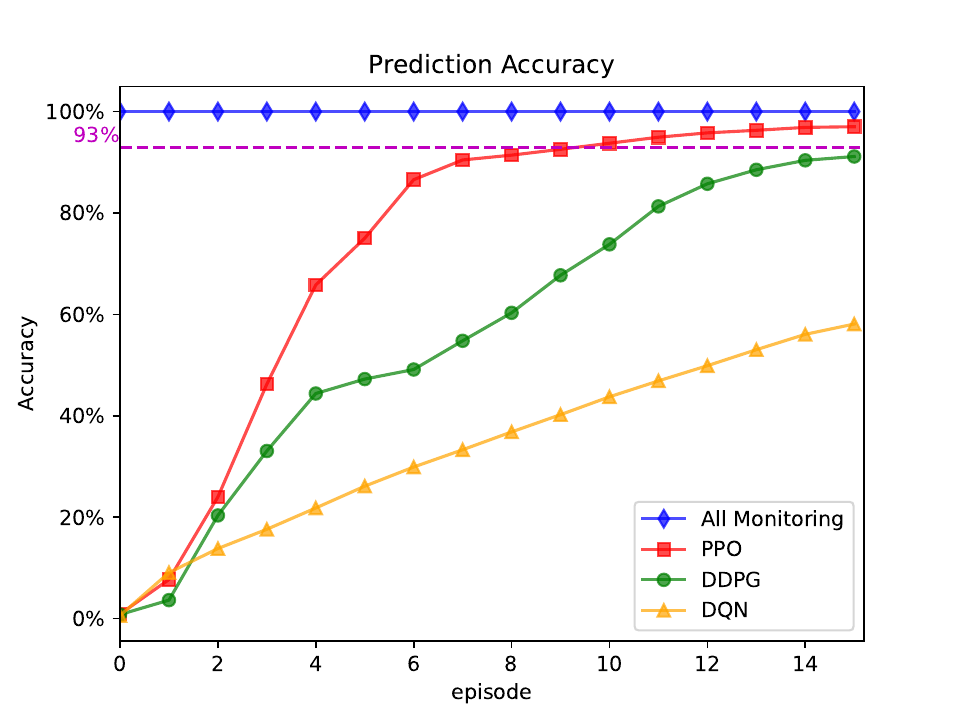}
		\caption{Comparison of different algorithms for prediction accuracy}
	\end{subfigure}
	\caption{Comparison of system performance among different algorithms}
	\label{fig3}
\end{figure}

As shown in Fig. 3(c), we compared the sleep duration of the three algorithms, and the DQN based strategy and the DDPG based strategy were less stable during the monitoring process. In comparison, the PPO algorithm shows more significant results. After 300 episodes, it was able to accurately predict the time of abnormality occurrence consistently, effectively implementing the proposed dynamic sleep strategy. Finally, the monitoring accuracy is shown in Fig. 3(d), where we output the ratio of the predicted number of anomalies to the actual number of anomalies per 100 events. It is worth noting that we stipulate that the monitoring accuracy rate must reach $93\%$ under the four strategies. As can be seen, there is no significant difference in the accuracy of the three algorithms until around episode 2. The DQN based strategy and the DDPG based strategy were unable to reach the required threshold within the specified 15 episodes. In contrast, the proposed PPO-based strategy can reach the threshold accuracy within 8 episodes, and the accuracy continues to increase as the learning process proceeds.
\section{Conclusion}
In this paper, we design an MAF dynamic dormancy mechanism based on the NWDAF framework within the NS-IoT system. The IoT devices subscribe to NWDAF service via MAFs, thereby acquiring the MAFs dormancy strategy, that is how monitoring and dormancy periods are allocated within defined time intervals. During an MAF dormancy phase, the devices under its supervision cease transmitting analysis data to the server, resulting in energy conservation. We model the dynamic monitoring duration adjustment process as an MDP process and propose the PPO algorithm, which is a kind of RL algorithm with superior convergence performance, to obtain the optimized MAF dormancy strategy. The experimental results show that our proposed algorithm outperforms other comparative algorithms in terms of energy consumption, dormancy duration, and monitoring accuracy.
\section*{Acknowledgments}
The work of this paper was supported by the National Natural Science Foundation of China(No.62171387), and the China Postdoctoral Science Foundation (No.2019M663475)
	\bibliographystyle{IEEEtran}
	\bibliography{sampleBibFile}
\end{document}